\documentstyle[12pt,epsf,psfig,color,lscape,mrs2003]{article}

\def\simlt{\lower.5ex\hbox{$\; \buildrel < \over \sim \;$}}
\def\simgt{\lower.5ex\hbox{$\; \buildrel > \over \sim \;$}}
\def\simpropto{\lower.2ex\hbox{$\; \buildrel \propto \over \sim \;$}}

\begin{document}
\def\simlt{\lower.5ex\hbox{$\; \buildrel < \over \sim \;$}}
\def\simgt{\lower.5ex\hbox{$\; \buildrel > \over \sim \;$}}
\def\simpropto{\lower.2ex\hbox{$\; \buildrel \propto \over \sim \;$}}

\title{SOME CURRENT ISSUES IN GALAXY FORMATION}
\author{Joseph  Silk}
\affil{Astrophysics,
Denys Wilkinson Building,
Keble Road,
Oxford OX1 3RH, UK}

\begin{abstract}
I describe recent challenges in hierarchical galaxy formation theory,
including the formation of disk galaxies and of ellipticals.  Problems with
cold dark matter are summarized, and possible solutions are presented. I
conclude with a description of the prospects for observing one of the most
important ingredients in galaxy formation theory, namely cold dark matter.
\end{abstract}


\section{\bf 
  Challenges of galaxy formation theory}

Galaxy formation theory must  account for  the properties and
evolution of galaxies, the  
star formation rate, the 
spectral energy distribution
and galaxy 
morphologies.  
Another important confrontation with observation
is with the
scaling relations. These  relations ({\it e.g.} Tully-Fisher,  fundamental
plane)
are controlled
by  the current relaxation time-scales (dynamical and  chemical) which  are long
 compared to the age
 of the universe. 
This is not an easy task because the theory is  almost entirely
phenomenological and is driven by the observations.
The ultimate 
 aim is to make  predictions at high redshift for  the current and future
generations of powerful detectors and very large telescopes.
Progress is inevitably iterative and slow, and observations are usually well
ahead of theory.
 A major hurdle is that there is no  fundamental  theory of
star formation.
Major uncertainties include the  initial stellar mass function,
 the  star formation efficiency and the 
 star formation rate. Of course,
 the empirical evidence for star
 formation is overwhelming, and this leaves cosmologists with little choice but to 
extract every possible output from their theories.

\section{ 
 Hierarchical galaxy formation}

The  ab initio approach to large-scale
structure has undergone a revolution in the past twenty years,
with an understanding of the initial conditions of structure formation.
Growth from inflation-boosted quantum fluctuations provides the current
paradigm that sets the point of departure for virtually all theories of
large-scale structure. The theory of structure growth  made one
notable prediction that has been verified with outstanding success.
This was the existence of 
fossil cosmic microwave background temperature fluctuations imprinted on the
last scattering surface of the cosmic microwave background.
The fluctuations are  on  angular
scales that correspond to the comoving scales of the observed large-scale
structure in the galaxy distribution. The WMAP satellite, adding 
unprecedented precision to 
many earlier experiments,
most notably those of BOOMERANG,  MAXIMA and DASI, has verified to within a
factor of order unity one of the most  remarkable  predictions of cosmology, 
confirming the
growth of structure via  the gravitational
instability of primordial density fluctuations

With the initial conditions specified, it became possible to simulate 
galaxy formation.
Three distinct approaches have emerged: numerical, semi-analytical and
hybrid. The fully numerical approach canot yet cope with the complexities of
star formation, but has been instrumental in guiding us towards an
understanding of the dark matter
distribution. The  semi-analytical approach has had most success, because
it can cope with a wide dynamic range via the extended Press-Schechter
formalism, to which is added a prescription for star formation based on
baryonic dissipation. The hybrid approach,
combining N-body simulations with a star formation prescription,
 is useful for its predictive power
in  observational cosmology, as it is ideal for constructing mock catalogues
of galaxies.

There have been some notable successes in the theory of  semi-analytic galaxy
formation. These  include an understanding of the 
large-scale clustering of galaxies via the primordial density fluctuation
power spectrum $P(k)$, including 
the two-point correlation function
$\xi(r)$ and its   higher moments, the predictions of the existence of 
filaments and sheets in the galaxy distribution and of the morphologies of
galaxy clusters,  the derivations of the
cluster and galaxy mass functions, and the predictions of 
large-scale velocity fields and weak lensing optical depths.
On smaller scales, the predictions of galaxy rotation curves
and of strong lensing  by massive galaxies and galaxy clusters
are generally considered to be successes of the
theory.
Global results that have motivated many  observations
which are in general agreement with the theory 
include the cosmic star formation history and the 
distribution  and evolution of HI clouds in the intergalactic medium
  $d^{2}N\over{dzdN_{HI}}$.

\subsection{\bf Disk  galaxy formation}

The rotation of galaxy disks
arises  from the generation of tidal torques between 
nonlinear density fluctuations in the process of forming protogalaxies.
The  angular momentum acquired by dark halos suffices to explain the
specific angular momentum of disks, if angular momentum is approximately
conserved  as the baryons collapse. Once the cold gas disk forms,
it is prevented from  prematurely forming stars because of gravitational 
self-regulation via the operation of large-scale disk gravitational
instabilities that are controlled by the cold gas fraction and by
the continuing supply of cold gas that depends on the star formation rate
and stellar feedback.
Self-regulation maintains the Toomre parameter, which is inversely
proportional to gas surface density and proportional to differential 
rotation-induced shearing motions  and to
turbulent velocity dispersion, to satisfy  $Q\sim 1$. One can visualise
the sequence of self-regulation as being:
$$ gas\  cools\ \Rightarrow \ Q \searrow\ \  \Rightarrow stars
\ form \ \Rightarrow heat\  gas\  \Rightarrow Q \nearrow .$$ 
 The empirical star formation rate can be written as 
$$ d\rho_\ast/dt=
\epsilon
\, \,  \mu_{gas}^n\, \, \Omega(r)\, \, f(Q) \  + \  
early \ \ gas \ \  infall.$$
Here $\epsilon$ is a measure of the efficiency of star formation,
$\Omega(r)$ is the disk rotation rate,  $n$ is empirically found to be
near unity, and $f(Q)$ $( \approx Q^{-2}-1   \, \,  { \rm if }\, \, 
Q<1,   { \rm but} \approx 0  \, \, { \rm if }\, \, 
Q>1 )$ is a threshold function that allows the outer disk
 to be
stable to star formation. In practice, this azimuthally-averaged 
description is too simplistic in the outer disk where 
 star formation is highly inhomogeneous. However 
an important  and generic consequence is that disks are expected to form inside-out.
Cold gas infall is required for chemical evolution, in order to account for
the paucity of metal--poor old disk stars (the so-called G dwarf  problem).
There is some 
evidence   from    recent HI observations that the high velocity HI
clouds around the Milky Way trace a more extensive gas reservoir
\cite{thil}, and may be manifestations of such cold infall.
The gas supply is expected to be  
disrupted in galaxy clusters. This is presumed to account for the
preponderance of early-type galaxies in clusters, and in particular the
dependence of morphological type on local density.

There are complications, however, that demonstrate that we have not yet
converged on the ultimate theory of galactic disk formation. The Tully-Fisher
relation is not understood. Models consistently give too high a normalisation
of mass at a given rotation velocity, due to the predominance of dark matter
in the model galaxies. There also is some question as to whether the slope is
well understood both for samples of nearby disk galaxies which have been carefully
corrected for inclination effects, and for distant disk galaxies
when projected forward in time for comparison with current epoch samples.
This may seem to be a detail, however the fundamental problems are twofold:
 most of the initial angular momentum in the theoretical models is lost to the
dark halo as the disk forms, and the distribution of observed angular momentum
is skewed towards high angular momentum in contrast to the initial
distribution predicted by the simulations.

The trigger that controls  star formation may be more complex than is inferred from the
Q parameter. There is evidence in our own neighbourhood for a series of 
multiple ministarbursts  rather than a monotonic decrease of star formation.
Yet another issue that is not well understood concerns the efficiency of 
star formation. This is the fraction of gas that is converted into stars
within a dynamical timescale. In our galaxy, the global value is about a
percent, thereby accounting for the longevity of star formation.
However surveys such as the Sloan Digital Sky Survey  are finding that
star formation efficiency increases as galaxy
mass increases \cite{kauf}. There does not seem to be a universal value for the 
star formation efficiency as might be expected if it were controlled only by 
the local physics associated with supernova feedback. 
Rather, global dynamical aspects must also play a role, for example via
regulation of disk stability. The stellar mass fraction increases relative to
the dark matter as the  disk mass increases. Massive disks are found to be
maximal, whereas less massive disks with smaller circular velocities are
usually submaximal \cite{kran}. The more extreme disk star formation efficiencies 
are inferred to have occurred in the
very early universe, simply in order to have  massive disks in place by a
redshift of unity \cite{cons}. Indeed,
at high redshift, such high star formation rates are
occasionally 
observed that the star formation efficiency must approach fifty percent or
more 
in systems undergoing extreme starbursts. These are presumably elliptical
galaxies in formation.

\subsection{\bf    Elliptical galaxy formation}

There is  no star formation theory for dynamically
hot systems such as elliptical galaxies. Appeal must be made to phenomenology.
Tidal interactions and mergers are found in simulations to be very effective
at concentrating
gas into the inner hundreds of parsecs.
Ultraluminous infrared galaxies are observed to have   star
formation 
rates of hundreds or even thousands of solar masses per year, as inferred if
the stars formed in a monolithic collapse of the system.
Post-starburst near-infrared light profiles are also
suggestive of forming spheroids. Since the ultraluminous infrared galaxies
are almost inevitably associated with ongoing mergers or strong tidal
interactions with nearby galaxies, it therefore seems entirely plausible
that these conditions are capable of
 driving intense  bursts of star formation at the prodigious star formation
rates that are observed.
Measurements of the molecular gas masses in several such systems at high
redshift demonstrate that a 
very high efficiency indeed of star formation is required, with  some
$10^{10}\rm M_\odot$ of stars being inferred to form in $10^7$ years \cite{bert}.

 Colours and spectra of elliptical galaxies at redshift of unity or beyond 
are suggestive of  a very
early formation epoch, at least for the stars \cite {dokk} if not for 
overall assembly. Additional
information comes from the observed ratios of $\alpha/Fe$ abundances. These
are enhanced
in massive ellipticals.
A  star formation duration of less than
a hundred million years is inferred in order
to
avoid excessive gas phase contamination by iron-producing SNIa ejecta that
would otherwise overdilute the observed $\alpha/Fe$ ratio that originates
from SNII 
ejecta \cite{mara}.
Of course, to argue that the stars formed
early and rapidly does not necessarily imply that the
galaxy was assembled monolithically when the stars  formed.
 Assembly could
have post-dated star formation, although HST  imaging with the ACS makes this
possibility increasingly unlikely.
The cosmic star formation history is likely to be
dominated by the precursors of today's ellipticals 
 at $z\simgt 2$. Of course such a  probe, which relies on galaxy surveys,  is 
rest frame UV flux-limited.
However the extragalactic diffuse background light from FIR to optical/UV
wavelengths 
provides a glimpse of all the star formation that ever occurred 
in the universe. Here it seems likely that forming dust-shrouded
ellipticals 
dominate the 
far infrared background 
above $400\mu$m \cite {devr}.

\subsection{\bf   Unresolved issues in galaxy formation theory}

One  of the greatest puzzles in galaxy formation theory
concerns the distribution of the dark matter.
The cold dark matter concentration is predicted from N-body simulations
to follow a density profile:
$$\rho=\frac{A}{r^{\gamma}(1+r/r_s)^{3-\gamma}}.$$
Here, $r_s$ is a scale factor that is incorporated into the concentration
parameter, $c\equiv r_v/r_s,$ where $r_v$ is either the virial radius or the
radius at an overdensity, spherically-averaged, of 200.
The profile slope parameter $\gamma$ is measured in high resolution N-body simulations
(\cite{fuku}, \cite{nava}) to be $\gamma\approx 1.2 \pm 0.3,$ and 
the normalisation parameter $A$ reflects the epoch of formation, typically defined to be
when  half of the present mass was at overdensity of 200.

Unfortunately, observations seem to be in mild disagreement with this
predicted profile \cite{merr}. A low CDM concentration is observed in low surface
brightness dwarf galaxies where the rotation curve is well measured. The
predicted dark matter cusp is not usually seen; the typical profile has a
soft core, although the interpretation is compounded by issues of disk
inclination, of the HI distribution which is usually used to measure the rotation
curve, and of the possible mismatch between baryon and CDM potential well
depths. The problem is still there in more massive galaxies.
In the Milky Way, a low concentration of nonbaryonic dark matter is
inferred, with the argument being made  that no more than
 10 percent of the total mass
interior to the solar circle can be non-baryonic. Theory predicts something
like 50 percent for a  CDM-dominated universe.
However the gravitational microlensing optical depth towards the bulge
of our  galaxy
is used to assess the stellar contribution to the inner rotation curve, and
this is uncertain by a factor  of $\sim 3.$ A low CDM concentration
has also been invoked to account for the observed
deficiency  of dark matter in intermediate luminosity
ellipticals 
 to $\sim$ 5 effective radii \cite{roma}, relative to the CDM
 predictions, in 
 contrast with the well
known evidence for dark matter from x-ray and strong lensing studies 
of luminous ellipticals.
Other objections to the hypothesis of cold dark matter dominance
abound but are of less concern.  For example, preferentially rapid formation of
the more massive galaxies, in apparent violation of the CDM mass
hierarchy evolution with redshift, massive halos forming later
and more slowly, is accomplished  by advocating strong
positive feedback, for which mechanisms exist.  The inference of cores
in
the density profiles of galaxy clusters from strong lensing studies is
avoided by allowing small deviations from axial  symmetry \cite{bart}.

Another issue is that of 
dark matter clumpiness. Large numbers of dwarf galaxy halos 
 are predicted at masses comparable to
those of the dwarf galaxies in the Local Group, exceeding the observed
numbers by an order of magnitude or more.
If these systems formed stars, they would be in gross disagreement with observations. If the angular momentum of the baryons is mostly
lost to the
dark halo as the baryons contract to form the disk, according to simulations,
then 
disk sizes of spiral galaxies are predicted to be smaller by about a factor
of 5  than is observed. 
The baryons are clumped and lose angular momentum as a consequence of dynamical
friction on the dark matter.

A related prediction is that of the galaxy luminosity function.
If the mass in  stars tracks that in  dark matter, far too many small galaxies are
predicted.
Too many massive galaxies are also predicted. This has been noted
both in isolated  groups of galaxies at the $L_\ast$ level
\cite{lake} and for the field luminosity function,
 where an excessive frequency of  super-$L_\ast$ 
galaxies is expected if a modern value for the initial baryon density is adopted
\cite{bens}. The 
problem 
 arises because the baryons
fall into the dark matter potential wells, cool and eventually form stars.
There are simply too many cold baryons.
If one begins with the baryon fraction predicted by primordial
nucleosynthesis of about 15 percent, one ends up with about twice as many
baryons as are seen even for the Milky Way galaxy.
This issue  has been aggravated by recent studies which show that many of
the accreting baryons enter the disk cold, without shocking to the virial
temperature \cite{deke}.
This appears to be  the dominant form of accretion both  for low mass galaxies and at high
redshift.

\section{\bf  Resurrecting CDM}

It would seem that cold dark matter has certain difficulties to overcome.
One approach is to tinker with the
particle physics by modifying the dark matter, 
for example by introducing self-interacting or fluid dark
matter. This approach  is not only non-compelling from the physics
perspective but it 
has also resulted in about as many new difficulties as it purports to resolve.
Another strategy  is to modify gravity. The less said about this the better:
it seems to this author that one should only modify the laws of fundamental
physics in the case of true desperation. We are not there yet. 

A more promising approach is via astrophysics.  The dark matter distribution
is inevitably modified by the impact of astrophysical processes. These
include dynamical feedback, such as via a massive, transient, rapidly rotating
bar. Such gaseous bars are expected to form in the course of a major merger
that preceded the first episode of star formation in the protogalaxy, and
later would settle into the galactic disk. Indeed up to half of spiral
galaxies have significant stellar bars. The initial tumbling of the bar is
slowed by dynamical friction on the dark matter. This provides a substantial
heat source that is capable of softening the CDM cusp into an isothermal
core \cite {holl}, but see \cite{atha} for an independent appraisal
of bar-halo angular momentum exchange. The converse consequence 
is that to explain 
the observed stellar bars that are generally in
rapid rotation, one needs either a deficiency of dark matter, less than 10
percent of the total mass within the region where the bar is observed, or to
argue that the observed bars are young. Cold gas infall to disks produces
cold stellar disks that can subsequently become bar-unstable \cite{comb}. The jury is still out on the
history and secular evolution of bars.

A more  radical astrophysical approach appeals to the formation of
supermassive black holes in the protogalaxy. These must have formed
contemporaneously with the oldest stars, as evidenced by the remarkable
correlation between spheroid velocity dispersion and  supermassive black hole
mass that extends over more than 3 orders of magnitude. Gas accretion onto the 
supermassive black hole is inevitable in the gas-rich protogalaxy, and
provokes  violent outflows. It is these outflows that are viewed 
in the spectra of quasars, the most luminous objects in the universe, and
which are powered by accretion onto supermassive black holes.
These massive outflows of baryons can  provoke efficient star formation and preferentially
expel  the low angular 
momentum gas. In this way, one might hope to understand why disks and more
generally galaxies are the sizes they are, why spheroids formed with great
efficiency, why half of the baryons have apparently been expelled 
from massive galaxies \cite{silk}, and why only high
angular momentum gas remains to form the disk. As for the impact on the  dark
matter, rapid loss of more than half the mass in the inner core of the galaxy
should  leave an impact by softening the dark matter profile.

\subsection{\bf The case for massive early winds}

There are a number of reasons for believing that massive winds played an
important role in galaxy formation.
Enrichment of the intracluster gas is observed to $\sim Z_\odot/3$.
This cannot be explained by current epoch star formation activity, or indeed
by past activity unless substantial mass ejection occurred.
Intracluster magnetic fields are observed at a level that is about 10 percent
of the typical galactic value. Ejection of magnetic flux
from galaxies early in the lifetime of the galaxies seems to be the most
plausible explanation. At a redshift of about 3, the
Lyman break galaxies are inferred to have  outflows to  velocities of $\sim
600\rm km/s.$ More indirectly, absorption against background quasars near
these galaxies has revealed evidence for
a  proximity effect on the intergalactic medium. This is in the form
of  a deficiency of 
HI that is observed  as  an increase in the transparency to  Ly$\alpha$ and possibly
 CIV absorption extending out to about $\sim 1$ Mpc from the Lyman break
galaxies \cite{adel}.

However numerical simulations of early supernova-driven winds
fail to find any evidence for substantial
gas ejection from luminous ($\sim L_\ast$)
galaxies \cite{spri}. 
One can ask what is wrong with the hydrodynamic simulations?
 Firstly, the simulations lack resolution.
Rayleigh-Taylor instabilities enhance
wind porosity and 
Kelvin-Helmholtz instabilities enhance
wind loading of the cold interstellar medium. Both effects are certain 
to occur and will enhance
the wind efficacity.
Secondly, the simulation initial conditions assume that the winds are driven
by supernovae produced by massive stars whose initial mass function is similar
to that found in the solar neighbourhood. This is a dangerous assumption,
given that we have no fundamental theory of the initial mass function, and
that conditions both in massive starbursts and in the early universe may be
quite different from anything sampled locally. A top-heavy  initial mass
function
is one way to boost the specific energy and momentum input by up to
an order of magnitude. It has been speculated that a top-heavy  initial mass
function is necessary to account for the high efficiencies of star formation
observed in certain very high redshift ultraluminous infrared sources.
This option  has also been invoked in order to account for the
surprisingly high redshift of reionisation found by the WMAP satellite \cite{whit}
and for the intracluster gas enrichment \cite{lars}.

Another  possibility is that some of the early supernovae may in fact be hypernovae.
A hypernova has up to $10^{53}$ ergs of kinetic energy. If one supernova in
10 at high redshift is in fact a hypernova, the specific energy input is
boosted by up to
an order of magnitude. The case for an enhanced hypernova fraction at high
redshift is based on the nucleosynthetic evidence from abundances measured
for the oldest stars in our halo. The enhancements of zinc and chromium and
deficiency of iron in these stars can be  explained in terms of
 hypernova yields. In hypernovae, the energy
output is boosted by infall of the inner rotating core onto a black hole,
 and the corresponding ejecta mass
cuts for precursors of $\sim 25 \rm M_\odot$ reflect
the observed abundance anomalies relative to standard supernova yields
\cite{nomo}.
Hypernovae are also a possible source of the r-process nuclear enhancement
seen in the oldest stars.

Finally, the ubiquitous AGN, as traced by the presence of supermassive black
holes that amount to $\sim 0.001$ of the spheroid mass, would
inevitably have been activated in the gas-rich protogalactic environment.
 The supermassive black hole
is presumed to achieve most of its growth by  gas accretion from a
circumnuclear disk. This would inevitably have been accompanied by
intense jets of relativistic plasma that provide a means of exerting strong
positive feedback onto the protogalactic environment \cite{breu}.

\subsection{\bf A new theory of outflows}

 Consider a multiphase interstellar medium heated by supernovae.
 The entrainment and porosity of the cold gas is  controlled by subgrid
 physics,
and one initially has to resort to a simple approach.
The galactic outflow rate can be written as 
$$\dot M_{outflow}=\beta\dot M_\ast f_{hot}.$$
This expression assumes that porosity, via the volume fraction of hot gas,
controls the  outflow rate of hot gas, which is also modulated by the mass of
entrained gas from the cold interstellar medium. 
 The  effective mass-loading  factor is defined by 
$$\beta=(1+L)\frac{\Delta m_{SN}}{m_{SN}}\sim 1,$$ 
where $L$ is the wind-loading factor, estimated from  Chandra observations of starbursts,
$\Delta m_{SN}$ is the mass ejected by a supernova and   
$m_{SN}$ is the mass
 consumed in star formation  in order for one Type II supernova to form.
The 
 hot gas filling factor is expressed in terms of the porosity as 
$f_{hot}=1-e^{-Q}$. 

I define the 
porosity $Q$ such that  
$ Q\equiv$ (SN  bubble  rate)$\times$(maximum  bubble  4-volume).
The porosity is  inferred to be
 proportional to the product of the rate of 
 star 
formation  and a factor $ p_{turb}^{-1.36},$ where 
I have assumed that the 
turbulent  gas pressure  is
$\rho_{gas}\sigma_{gas}^2.$
Next,  I rewrite the porosity in terms of the
star formation rate as $$Q=\frac{\dot\rho_\ast}{G^{1/2}\rho_{gas}^{3/2}}
\left(\frac{\sigma_f}{\sigma_{gas}}\right)^{2.7},$$
where the fiducial velocity $\sigma_f$ absorbs the constant of proportionality.
Now define the supernova explosion kinetic energy $E_{SN}.$ 
Typical values of $E_{SN}$ and  $m_{SN}$ are $10^{51}$ ergs and $300\rm
 M_\odot$, respectively, for a normal initial stellar mass function, similar
 to that measured locally.
One can then  write  the fiducial velocity  as 
$$\sigma_f\approx 20 \rm km s^{-1}\left(E_{SN}/10^{51}\right)^{1.27} 
\left(200\rm M_{\odot}/m_{SN}\right).$$  This determines the effectiveness of
porosity relative to some specified turbulence field maintained by the galaxy
potential well. Recent numerical simulations by A. Slyz verify that the star formation rate in
a multiphase star-forming medium is indeed given by 
$\dot\rho_\ast \propto Q p_{turb}^{1.5}.$ Since the porosity tends to
self-regulate, this means that there is positive feedback, resulting in a
burst of star formation as pressure from supernova energy injection
builds
up.
The process stops when the gas supply is exhausted.

I finally  write the star formation rate as
$\dot \rho_\ast=\epsilon \rho_{gas}\Omega$. The dimensionless star formation 
efficiency $\epsilon$ incorporates the hidden physics of porosity and
turbulence: it is not a constant. Indeed the SDSS survey of 100,000 galaxy
spectra
finds that star formation efficiency increases with galactic stellar mass
up to
a mass of about $3\times 10^{10}\rm M_\odot,$ above which it levels off.
A similar result is inferred from the present analysis, if the appropriate
physics is incorporated into the fiducial parameter $\sigma_f.$ One can see
this
by writing the porosity as 
$$Q=\epsilon\left(\frac{\rho}{\rho_{gas}}\right)^{1/2}\left(\frac{\sigma_f}{\sigma_{turb}}\right)^{2.7}.$$
If the porosity self-regulates, the star formation efficiency increases
approximately as $\sigma_{turb}^3.$
There is a critical value of $\sigma_{turb},$ below which the porosity is large. 
 If  the porosity is large, $ Q\simgt 1,$ and 
$\dot M_{outflow}\approx\dot M_\ast$. The outflow rate is of order the star
formation rate. This is a generic result, and is similar to
 what is observed
in nearby starburst galaxies, which are generally  low mass galaxies. I
speculate that 
 $\sigma_f$ is likely to  be  boosted relative to the preceding simplistic
analysis by some combination of the following: instability-enhanced porosity,
hypernovae,
a top-heavy IMF and AGN-triggered outflows.
The model explains the observed starburst superwinds and
the observed star formation efficiency dependence  on galactic stellar
mass if $\sigma_f \approx 100 \rm km/s.$ At early epochs,  $\sigma_f$  is likely to be
larger,  and such outflows should also prevail in massive galaxies.

\section{\bf  Observing CDM:  motivated candidate is WIMP LSP}

 Assuming that the WIMPS once were in thermal equilibrium, 
one finds that the relic WIMP froze  out at
$$n_x<\sigma_{ann}v>t_H\simlt 1\Longrightarrow T\simlt m_\chi/20k.$$
From this, one infers that the 
relic CDM density is $\Omega_x\sim 
\sigma_{weak}/\sigma_{ann}$. It is useful to know the mass range of the WIMPs
in order to define  search parameters. Minimal SUSY has many free parameters,
and most of them are generally  suppressed.  For example, requiring the relic neutralino
density to be within  mSUGRA  greatly reduces the parameter space for
possible masses \cite{edsj}.
If the WIMP is a SUSY  neutralino, simple scaling arguments yield 
$$<\sigma_{ann}v> \propto m_\chi^2 \ \ {\rm  for} \ \ 
m_\chi\ll Z^0$$
and 
$$<\sigma_{ann}v> \propto m_\chi^{-2}  \ \ {\rm for} \ \ 
m_\chi\gg Z^0,$$
thereby defining a window of opportunity for dark matter.
Stability is assumed for the SUSY LSP to be a WIMP candidate, usually via
R-parity conservation.
From accelerator limits
combined with  model expectations,
the allowed mass range is conservatively found to satisfy
  $$ 50 {\rm  GeV} \simlt
m_\chi  \simlt 1 {\rm TeV}.$$ 
Accelerator limits set a lower bound, and 
the inclusion of the extra degrees of freedom from coannihilations sets
 an upper bound.
Direct searches may also independently set a model-dependent  lower bound.

Indirect searches via halo annihilations of  the LSP into $\gamma,
\bar p,
e^+,\nu $ have hitherto been inconclusive. 
There are hints of an anomalous feature in the high energy $e^+$
spectrum. However 
halo detection of $e^+$ requires
clumpiness of order
$$<n^2>/<n>^2\sim 100,$$ both to get sufficient flux and to allow the
possibility of a nearby clump which might allow the observed spectral
feature to be reproduced \cite{hoop}. Such clumpiness could also boost the predicted
gamma ray flux from annihilations into the range observable by EGRET.
Clumpiness of this order is indeed predicted by galaxy halo simulations.
However this generally applies in the outer halo. The 
 $\gamma$-ray flux towards the galactic
centre  is observed to have a  hard 
spectrum (as expected for   annihilations), but the clumps would not survive
the
tidal disruptions that are inevitable in the inner galaxy \cite{stoe}.
To account for  the observed diffuse gamma ray flux from the direction of the 
 galactic
centre, one
 would need to have a very steep density profile
($\rho\simpropto r^{-1.5}$). This would conflict with microlensing
observations and the inner rotation curve of the Galaxy.

\section{\bf The future}

There are exciting prospects for addressing many of the challenges facing
galaxy formation and  dark matter.
With regard to directly observing 
 forming galaxies, we can look forward to sampling the galaxy luminosity
function
at  redshifts beyond unity  with both  
 SIRTF and ground-based NIR spectroscopy.
The theory of  multiphase galaxy formation is certain to be greatly refined,
incorporating 
dynamical feedback and the impact of supermassive black holes.
We will probe scales down to  $\sim 10^6\rm M_\odot$ via 
spectroscopic gravitational lensing. Baryonic dark matter 
 will be mapped at  UV/SXR wavelengths.
In the area of indirect detection of CDM, new experiments will  search for 
high energy halo annihilation signatures in the form of 
$\gamma, e^+, \bar p$ and   $\nu$.
Over the next 5 years, these experiments will include
 GLAST, HESS, MAGIC, VERITAS, ICECUBE, ANTARES, PAMELA and AMS. High energy neutrinos
from annihilations in the  sun (and
earth) will be probed, thereby providing a measure of the  cold dark matter
density at the  solar circle.

 The Galactic Centre could provide a ``smoking
gun" with radio synchrotron, $\gamma$-ray and $\nu$ data: 
annihilations  measure cold dark matter  where
Milky Way formation began ``inside-out", some 12 Gyr ago.
Accretion models onto the central black hole fail to give  sufficient low
frequency radio or gamma ray emission to account for the observed fluxes from
SagA$^\ast$,
and it is tempting to invoke a more exotic alternative.
However even if the history of the supermassive black hole  
at the centre of the galaxy disfavours a  cold dark matter spike as has been
argued \cite{merrit},
one might expect  lesser spikes to survive around other relic  masssive black
holes. The central supermassive black hole and the bulge of the galaxy most likely formed 
from the mergers of protogalactic dwarf
galaxies  that themselves contained smaller black holes. This model
suggests that   there should be 
relic ``naked" intermediate mass black holes in the inner halo \cite{isla}.
The adiabatic growth of these seed  black holes  should have generated local  spikes
in cold dark matter that could have survived and maintained 
 a density profile
$$\rho\propto r^{-\gamma}\Rightarrow\rho\propto
 r^{-\gamma^\prime}, \ \ {\rm  with} \ \   \gamma^\prime=\frac{9-2\gamma}{4-\gamma}.$$
Annihilation fluxes would be  enhanced, to a level where such sources could
possibly account for a subset of the unidentified EGRET gamma ray sources.

Of course, the preceding interpretation rests heavily on the hypothesis that
the dark matter consists primarily of the lightest $N=1$ SUSY neutralinos.
This is well motivated, but as has often been emphasized, the most compelling
and elegant explanation of any natural phenomenon is often false. Of course,
if accelerator evidence were found for SUSY, the odds in favour of a
neutralino explanation of dark matter would be dramatically increased.
   It is exceedingly difficult to construct a theory of galaxy formation without
some compelling evidence for the nature of the dark matter. We assume that
the dark matter is cold and stable, and this results in beautiful simulations
of cosmic structure that meet  many, but by no means all, of the
observational challenges. Our hope is that with increasingly refined probes
of galaxies near and far, we will be able to construct a strong
inferential case for the required properties of the dark matter.
Indeed, even now we are not far from this goal  in so far as our modelling of
large-scale structure is concerned. 

On smaller scales, however, the picture,
and the corresponding role of dark matter, is much less clear.
It is particularly disconcerting that we know so little
about  the fundamental physics of star formation, despites decades of 
detailed observations. It is only too tempting to assume that  conditions
in the distant universe, while being far more extreme than those encountered
locally, nevertheless permit us to
 adopt similar rules and inputs for star formation. We may be easily misled.
Galaxy formation moreover rests on knowledge of the initial conditions that
seeded structure formation, and that
we  measure  in the cosmic microwave background. 
Here too it is worth recalling that our conclusions are only as robust as the
initial priors. Change these substantially, and new modes of fluctuations are
allowed that can, for example, permit a
much earlier epoch of massive galaxy formation than in
the standard model. It is clear that    only 
increasingly refined and precise
  observations      will guide
us: if      evidence were to be confirmed for
a hypothesis that was far from our current prejudices, theory
     would rapidly adapt.
 We should bear in mind that
Nature has more surprises than we can imagine, otherwise physics would be
hopelessly dull.

\acknowledgements

I thank my colleagues especially at Oxford for their unwavering enthusiasm
in discussions about many of the topics covered here. In particular I
acknowledge the contributions of Celine Boehm, Greg Bryan, Julien Devriendt, Ignacio
Ferreras, Dan Hooper, Hugues Mathis,
Adrianne Slyz and James Taylor. In preparing this review, I deliberately restricted myself to
citing papers from 2003: apologies to those whose work has not been cited.

\end{document}